\documentclass[11pt]{article}
\usepackage{moriond,epsfig,amsmath}

\def\Journal#1#2#3#4{{#1} {\bf #2}, #3 (#4)}

\def\NPB{{\em Nucl. Phys.} B}
\def\PLB{{\em Phys. Lett.}  B}

\def\PRD{{\em Phys. Rev.} D}

\def\EJC{{\em Eur. Phys. J.} C}

\def\ra{\rightarrow}

\def\be{\begin{equation}}
\def\ee{\end{equation}}
\def\bea{\begin{eqnarray}}
\def\eea{\end{eqnarray}}

\begin{document}
\vspace*{4cm}
\title{BARYON PAIR, $\rho^0\rho^0$ PAIR AND INCLUSIVE HADRON PRODUCTION 
  IN TWO-PHOTON COLLISIONS AT LEP
\footnote{Talk presented at the XXXVIIIth Rencontre de Moriond (QCD), Les
Arcs, March 2003}}

\author{ C.H. LIN }

\address{Department of Physics, National Central University,\\
Jung-Li, TAIWAN 320}

\maketitle\abstracts{The most recent results of inclusive hadron, 
baryon pair and $\rho^0\rho^0$ productions in two-photon collisions measured at 
LEP are presented.
}

\section{Introduction}
Electron-positron colliders are a suitable place for the study of two-photon
interactions via the process $e^+e^- \ra e^+e^- \gamma\gamma \ra e^+e^-X$.
The outgoing electron and positron carry almost the full beam energy and are 
usually undetected, due to their small transverse momenta. When the scattered 
electron is detected (``tagged'') by the forward detector, an off-shell photon
$\gamma^*$ with a large squared four-momentum, $Q^2$, is emitted. The final
state $X$ can be leptonic or hadronic. The cross-sections of two-photon
interactions with different 
final states are calculable by QCD or QED. In this 
report, we present some results of measurements of pair production and inclusive hadron 
production in two-photon collisions at LEP.

\section{Baryon pair production}
\begin{figure}
  \begin{center}
    \vskip 0.5cm
    \psfig{figure=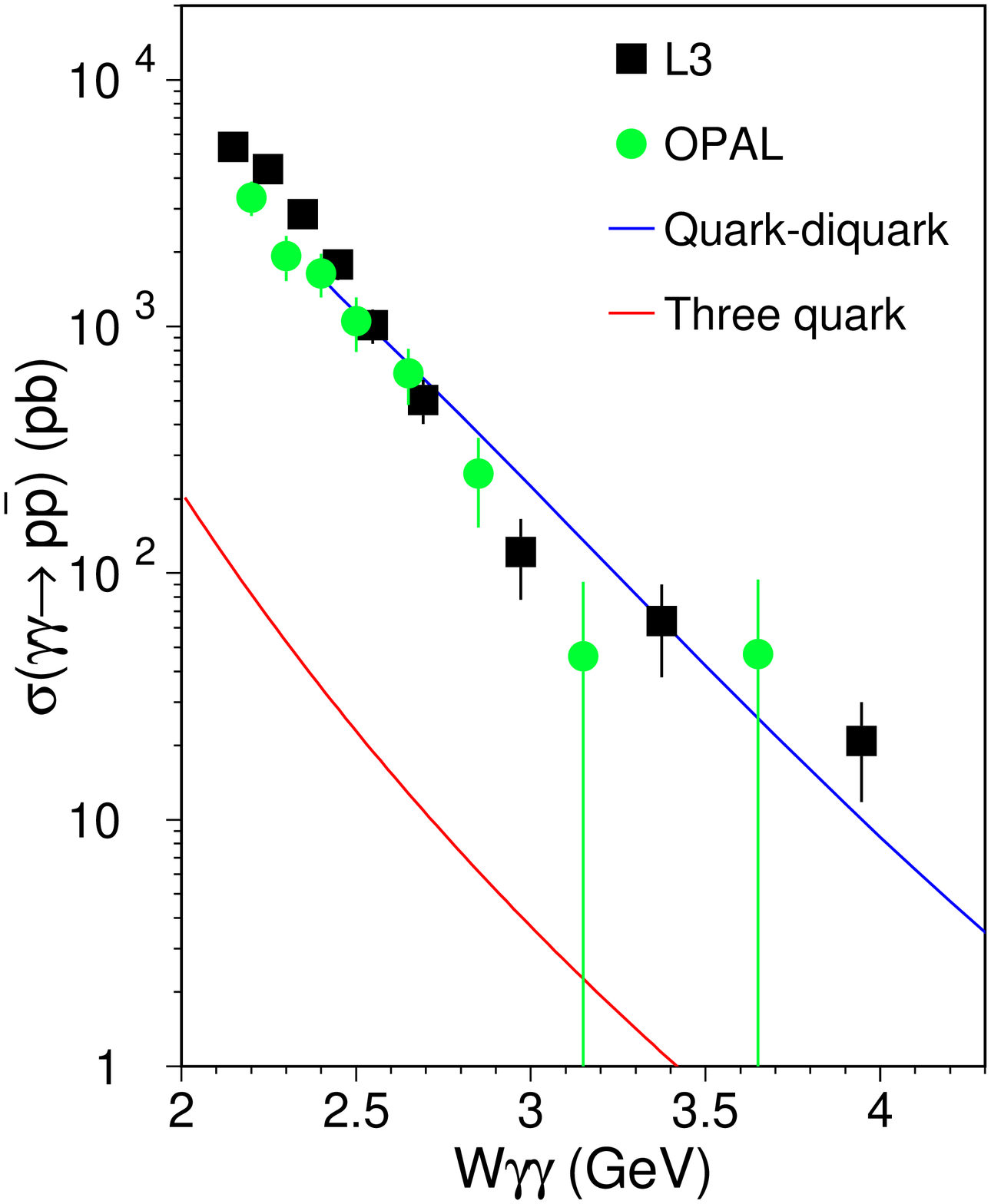,height=2.2in, width=1.85in}
    \psfig{figure=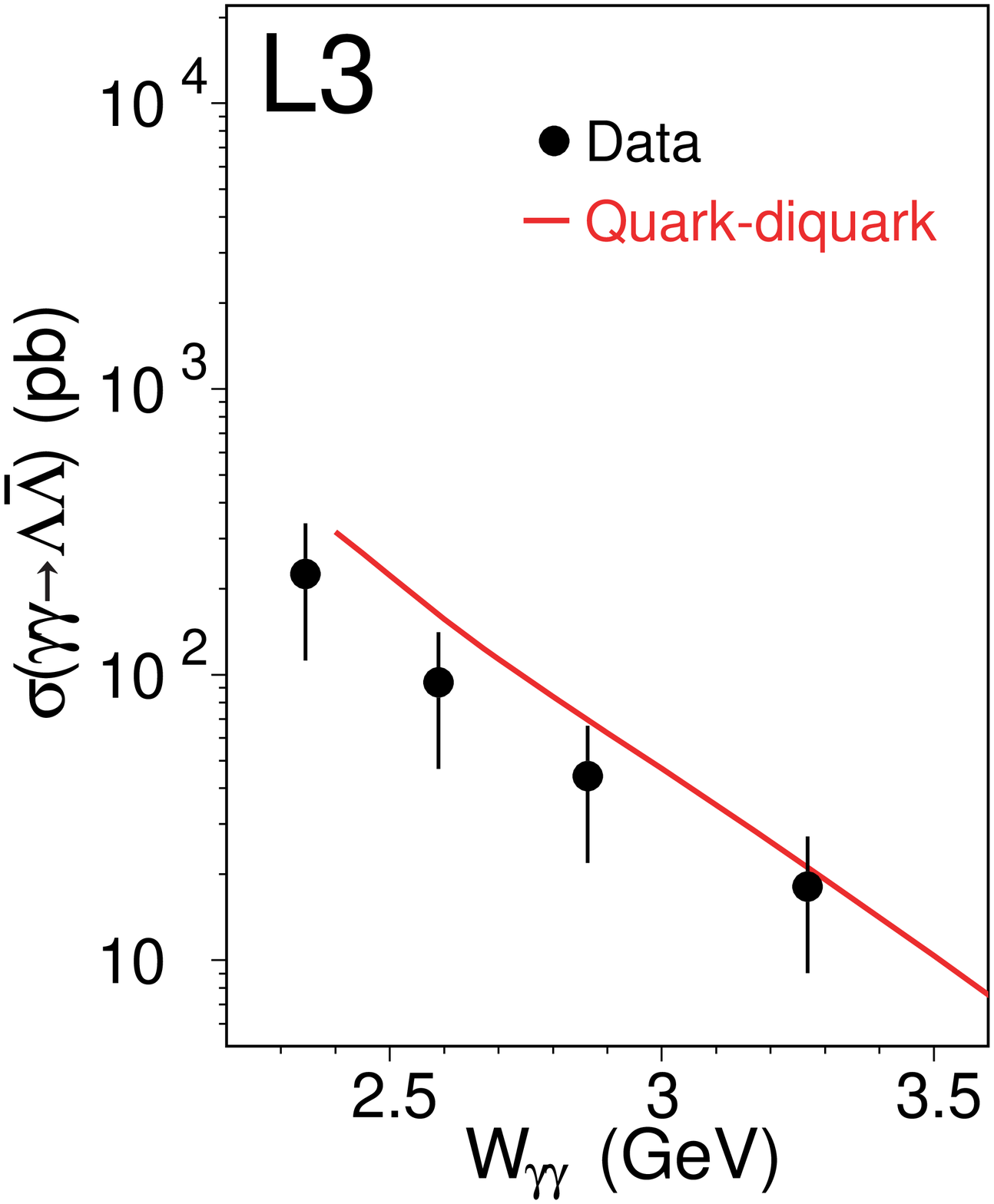,height=2.2in, width=2in}
    \psfig{figure=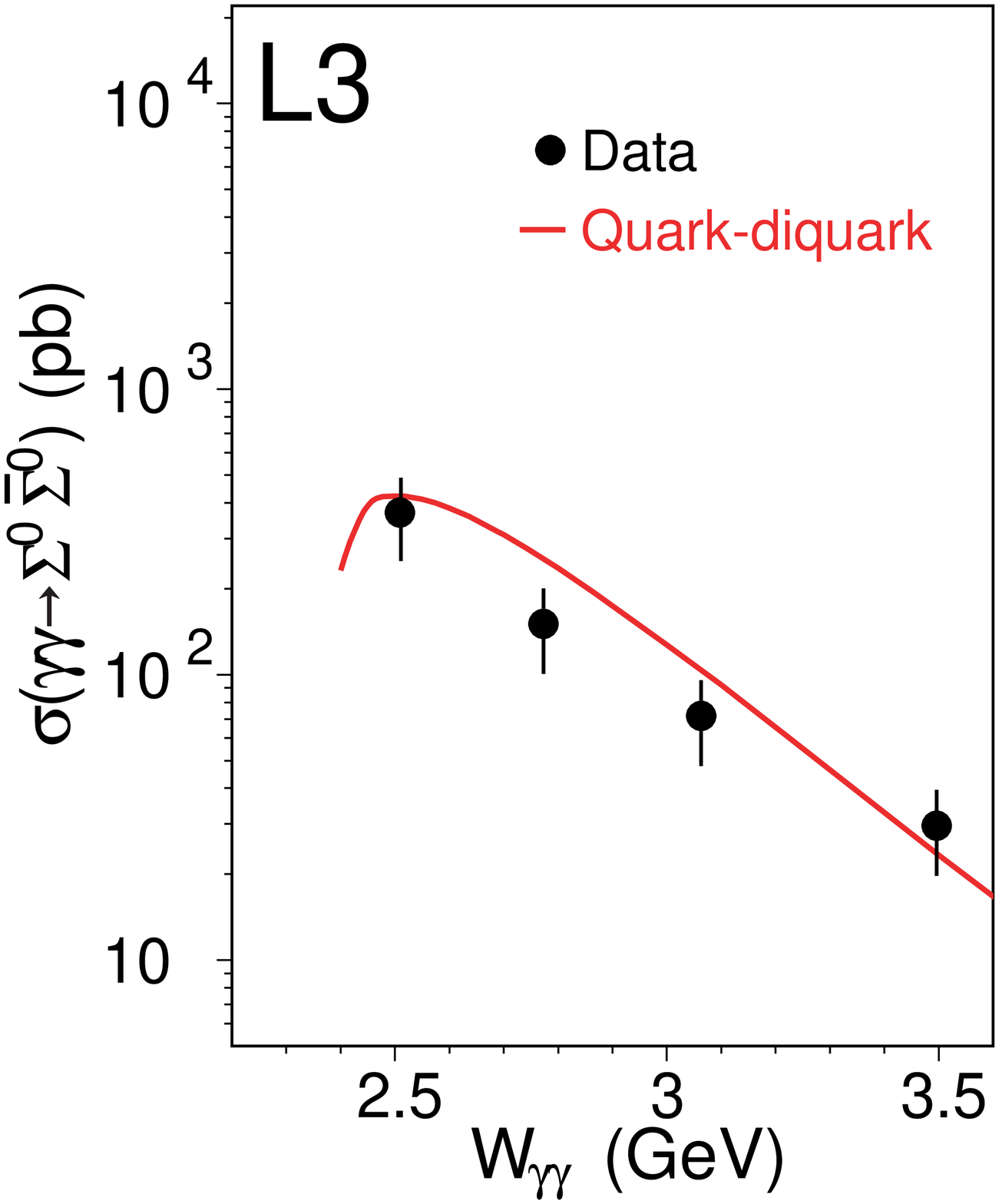,height=2.2in, width=2.1in}
    \caption{Cross-sections of $\gamma\gamma \ra p \bar{p}$, 
             $\gamma\gamma \ra \Lambda \bar{\Lambda}$ and
             $\gamma\gamma \ra \Sigma^0 \bar{\Sigma^0}$ processes.
      \label{fig:ppcros}}
  \end{center}
\end{figure}

\begin{figure}
  \begin{center}
    \vskip 0.5cm
    \psfig{figure=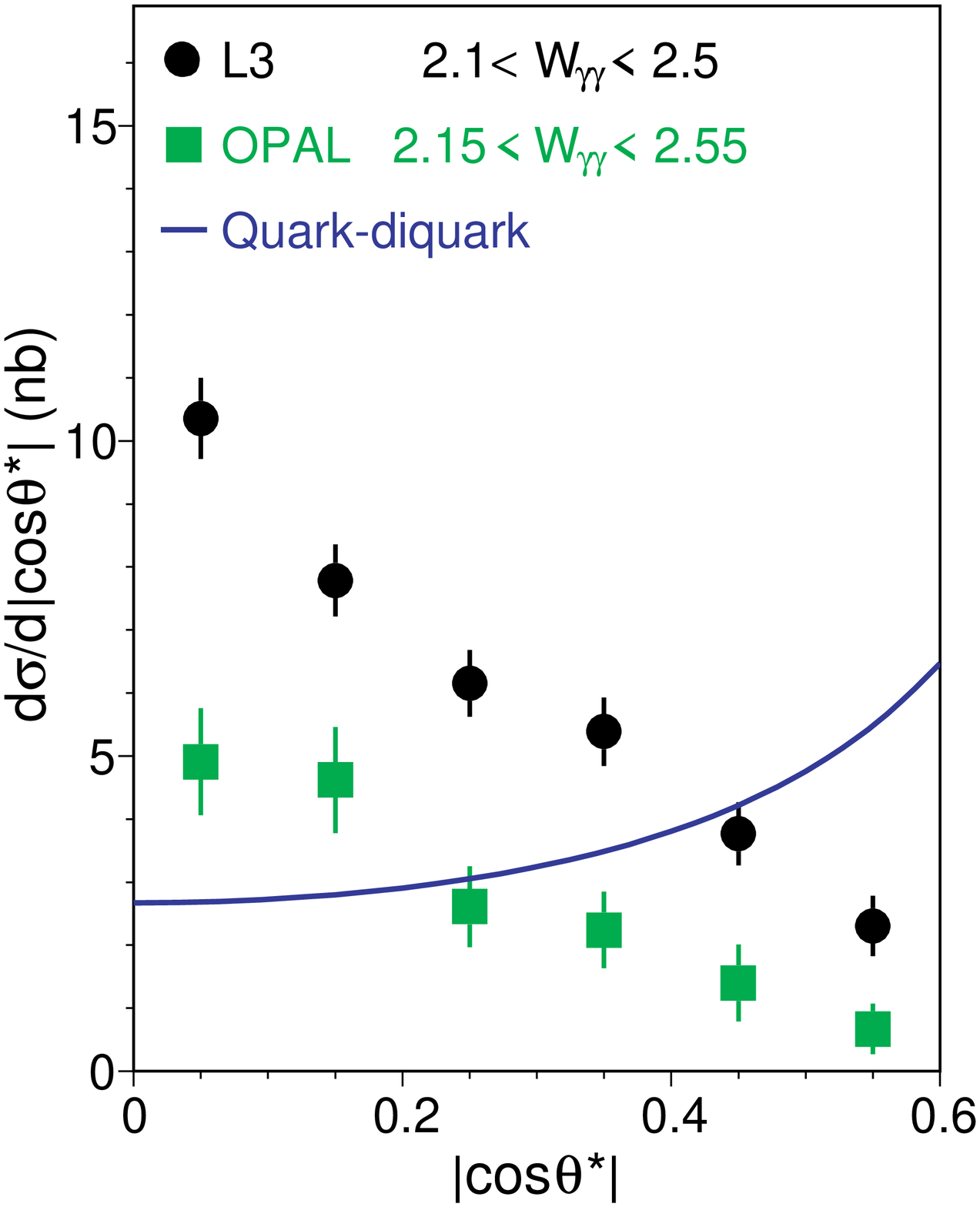,height=2.2in, width=2in}
    \psfig{figure=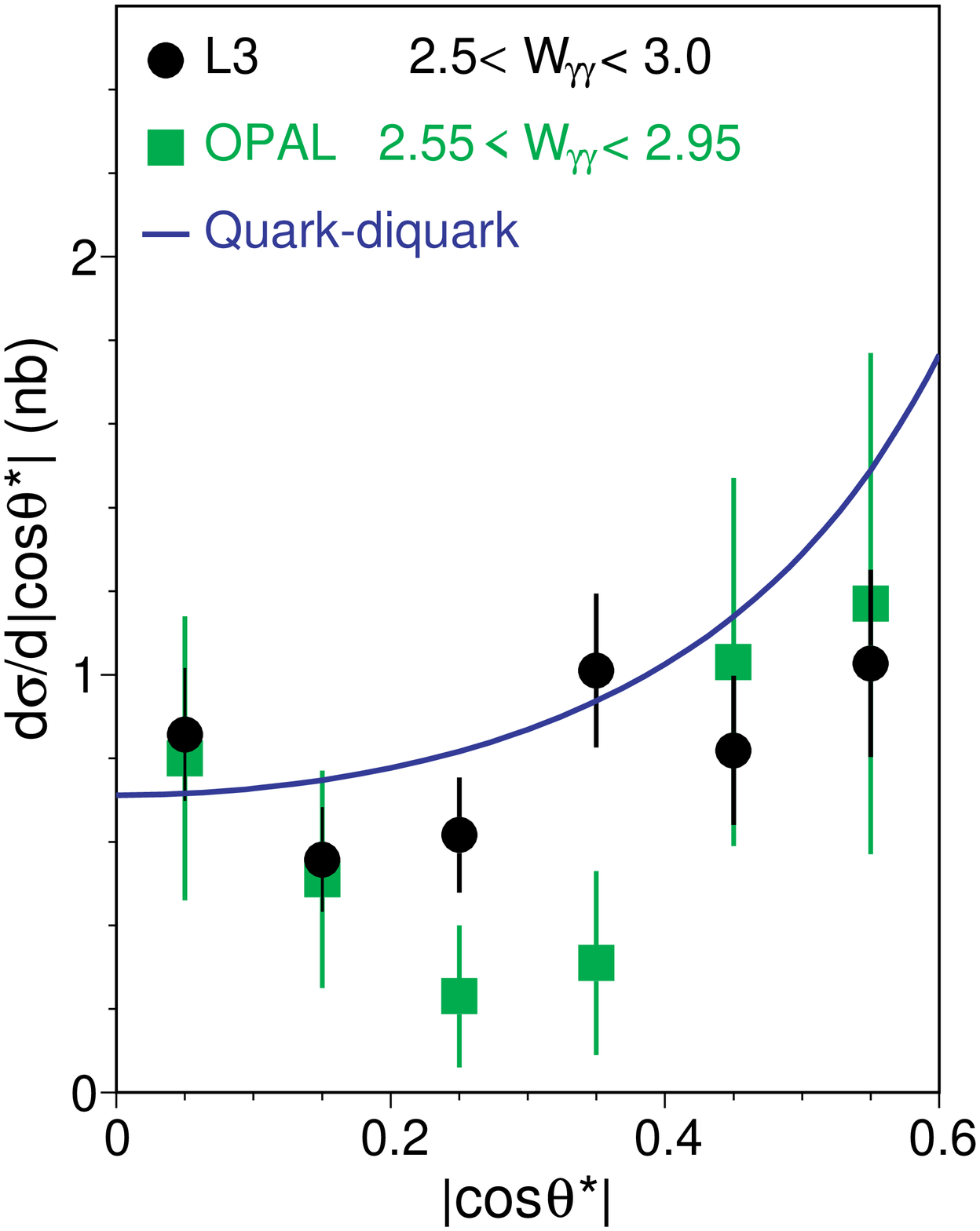,height=2.2in, width=2in}
    \psfig{figure=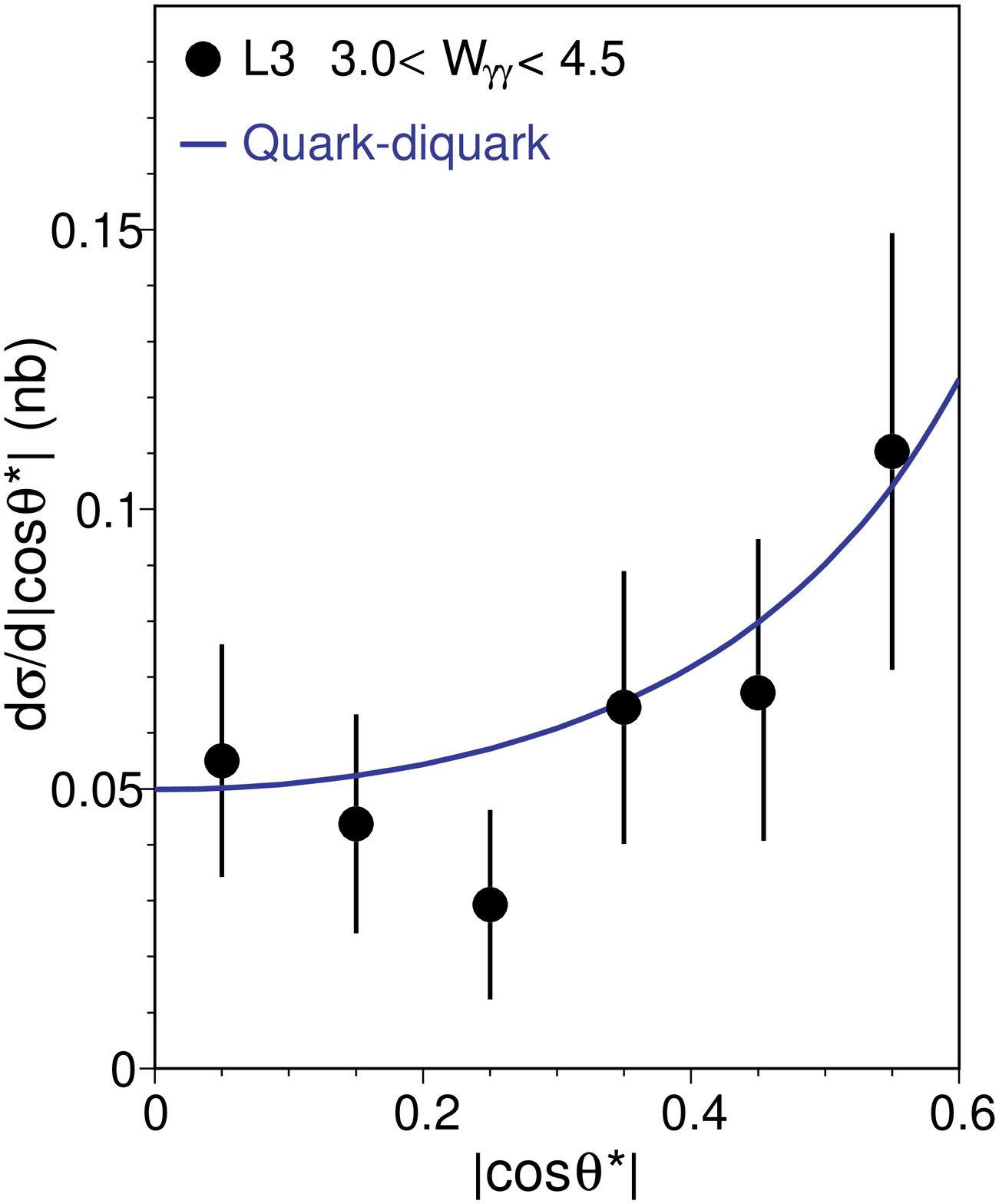,height=2.2in, width=2in}
    \caption{Differential cross-sections of $\gamma\gamma \ra p \bar{p}$
             as a function of $\cos \theta^*$ in three different 
	     $W_{\gamma\gamma}$ bins.
      \label{fig:ppcos}}
  \end{center}
\end{figure}

The process of $\gamma\gamma \ra baryon~antibaryon$ is sensitive to the quark
structure of the baryon, the cross-section of this process being calculable in the
framework of the hard scattering approach~\cite{hs}. At LEP three different 
processes of the baryon pair production were studied at $\sqrt{s} = 
183-209~\mathrm{GeV}$~\cite{baryon}:
\bea
    \gamma\gamma & \ra & p \bar{p} \nonumber \\
    \gamma\gamma & \ra & \Lambda \bar{\Lambda} \ra p \pi^- \bar{p} \pi^+ 
    \nonumber \\
    \gamma\gamma & \ra & \Sigma^0 \bar{\Sigma^0} \ra
    \Lambda \gamma \bar{\Lambda} \gamma \ra p \pi^- \gamma ~ \bar{p} \pi^+ \gamma
    \nonumber
\eea
In the study, protons and antiprotons were mainly identified by the energy 
loss and the ratio of energy and momentum measured by the central tracker and the
electromagnetic calorimeter. The measured cross-sections of the three processes
as a function of two-photon mass, $W_{\gamma\gamma}$, are shown in 
Figure~\ref{fig:ppcros}.
Good agreement is found between the L3 and OPAL experiments. The predictions of the
quark-diquark model~\cite{qq} are consistent with the data. The predictions of 
three-quark model~\cite{qqq} are too low. For the high statistics  
$\gamma\gamma \ra p \bar{p}$ process, its cross-section is also measured 
as a function of $\cos \theta^*$ in three $W_{\gamma\gamma}$ bins shown in
Figure~\ref{fig:ppcos}. In the low mass region the angular distribution is
strongly peaked at large angles and the quark-diquark models fails to describe
the data. The agreement improves in the high mass region, where the angular
distribution is instead peaked at small angles.

\section{Exclusive $\rho^0\rho^0$ production at high $Q^2$}
Exclusive $\rho^0\rho^0$ production with a highly virtual photon in 
two-photon collisions can be used to verify the mechanism
of $qq,~gg \ra \mathrm{meson~pair}$. 
Recently, perturbative QCD predictions~\cite{gda} of the cross-section
of such process has been made. 

\begin{figure}
  \begin{center}
    \vskip 0.5cm
    \psfig{figure=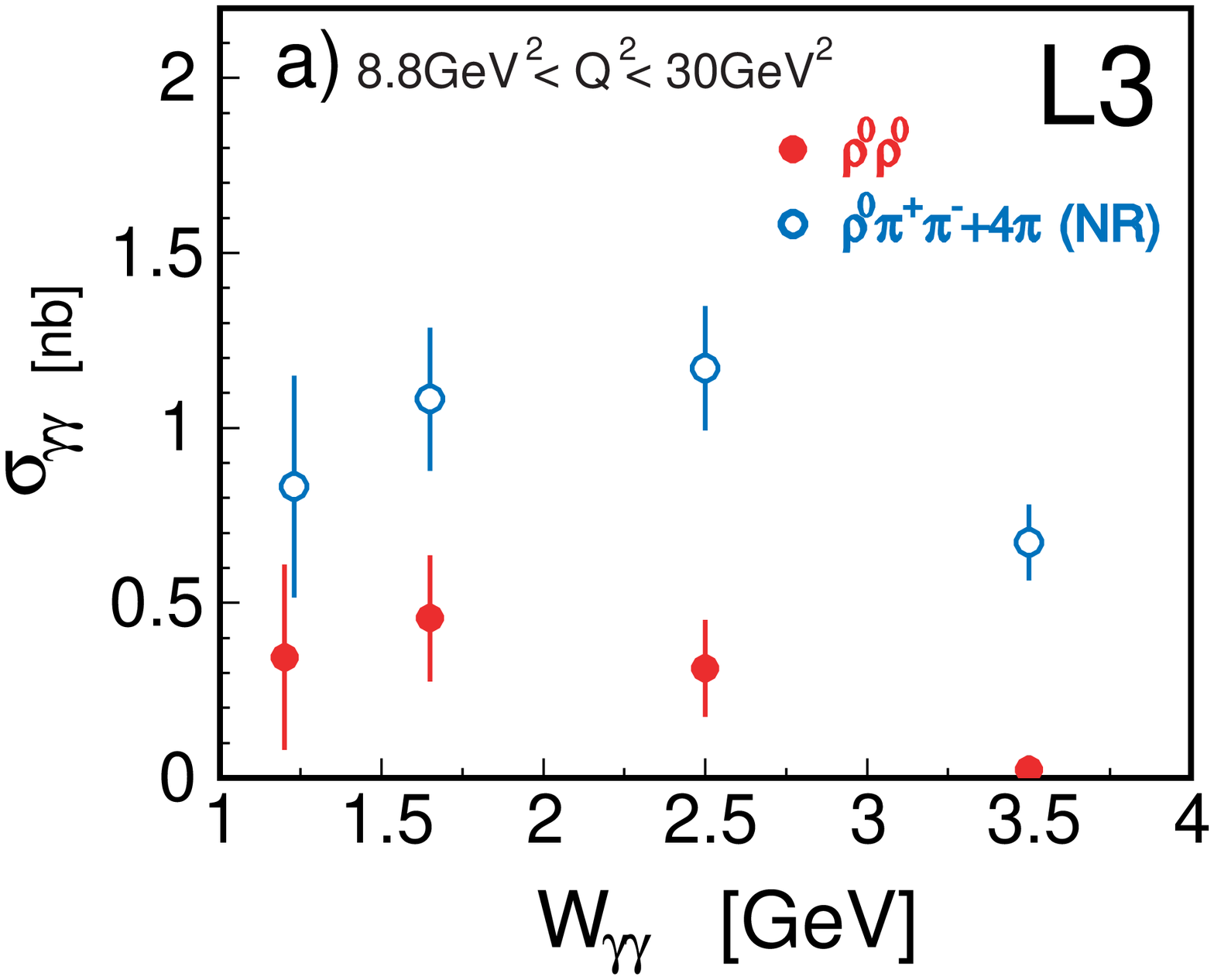,height=2.2in, width=2.2in} \hskip 0.5cm
    \psfig{figure=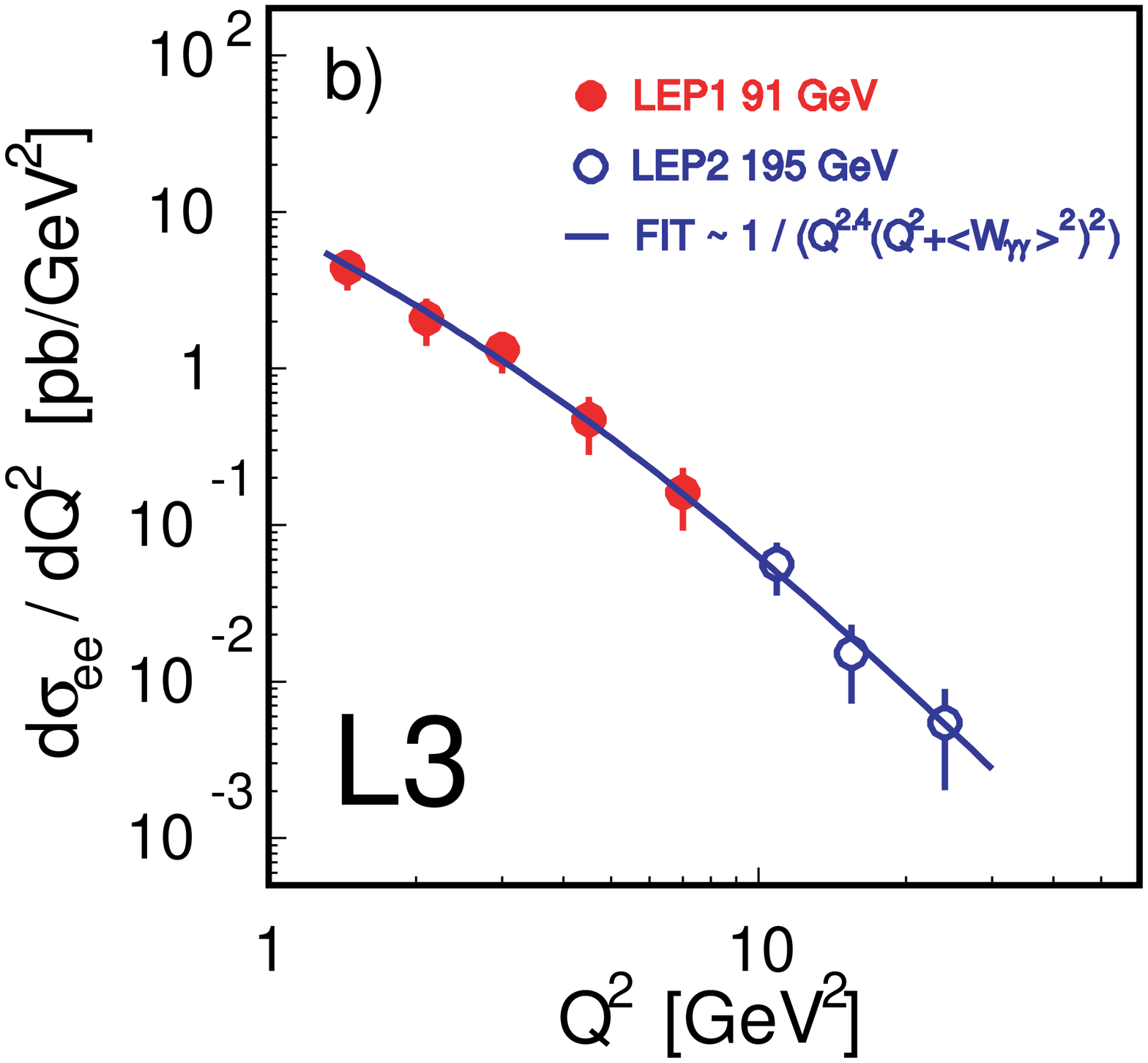,height=2.2in, width=2.3in}
    \caption{(a) Cross-section of the process $\gamma\gamma^* \ra \rho^0 \rho^0$
                 as a function of $W_{\gamma\gamma}$. (b) differential 
                 cross-section of the process $e^+e^- \ra e^+e^- \rho^0\rho^0$
                 as a function of $Q^2$.
      \label{fig:rhorho}}
  \end{center}
\end{figure}

The data used in this study were collected by the L3 detector at $\sqrt{s} =
89-209~\mathrm{GeV}$~\cite{rho}. The events were selected by identifying a
scattered electron and four charged pions in the detector. The background
processes $\gamma\gamma^* \ra \rho^0 \pi^+\pi^-,~ 4\pi(non-resonant)$ are
separated by a box method~\cite{rho}. Figure~\ref{fig:rhorho}a shows the
cross-section of the process $\gamma\gamma^* \ra \rho^0 \rho^0$ as a function of
$W_{\gamma\gamma}$. A board enhancement at threshold is observed. The
differential cross-section of the process $e^+e^- \ra e^+e^- \rho^0 \rho^0$ as
a function of $Q^2$, which is consistent
with the pQCD expectation, is shown in Figure~\ref{fig:rhorho}b

\section{Inclusive hadron production}
\begin{figure}
  \begin{center}
    \vskip 0.5cm
    \psfig{figure=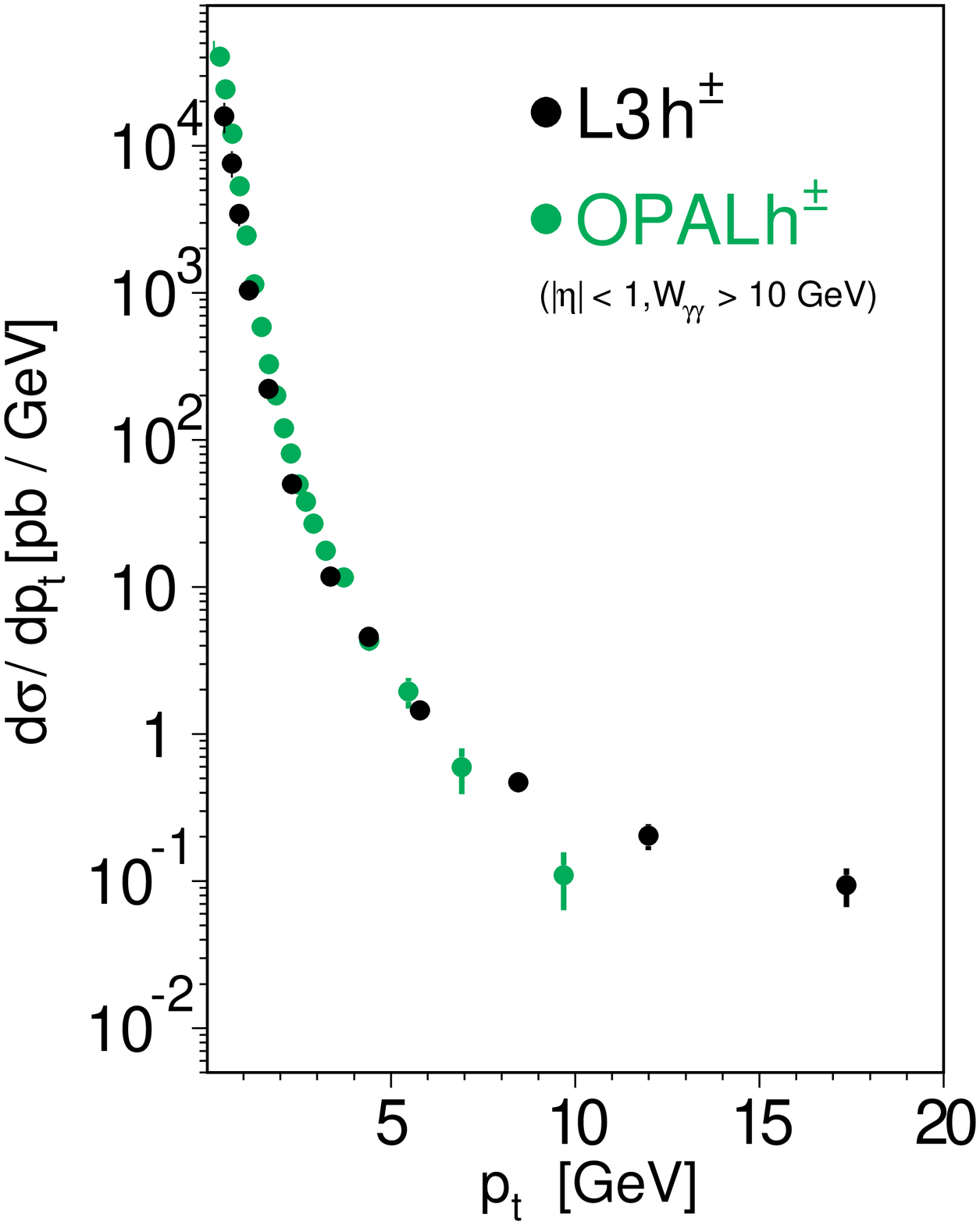,height=2.2in, width=2.1in}
    \psfig{figure=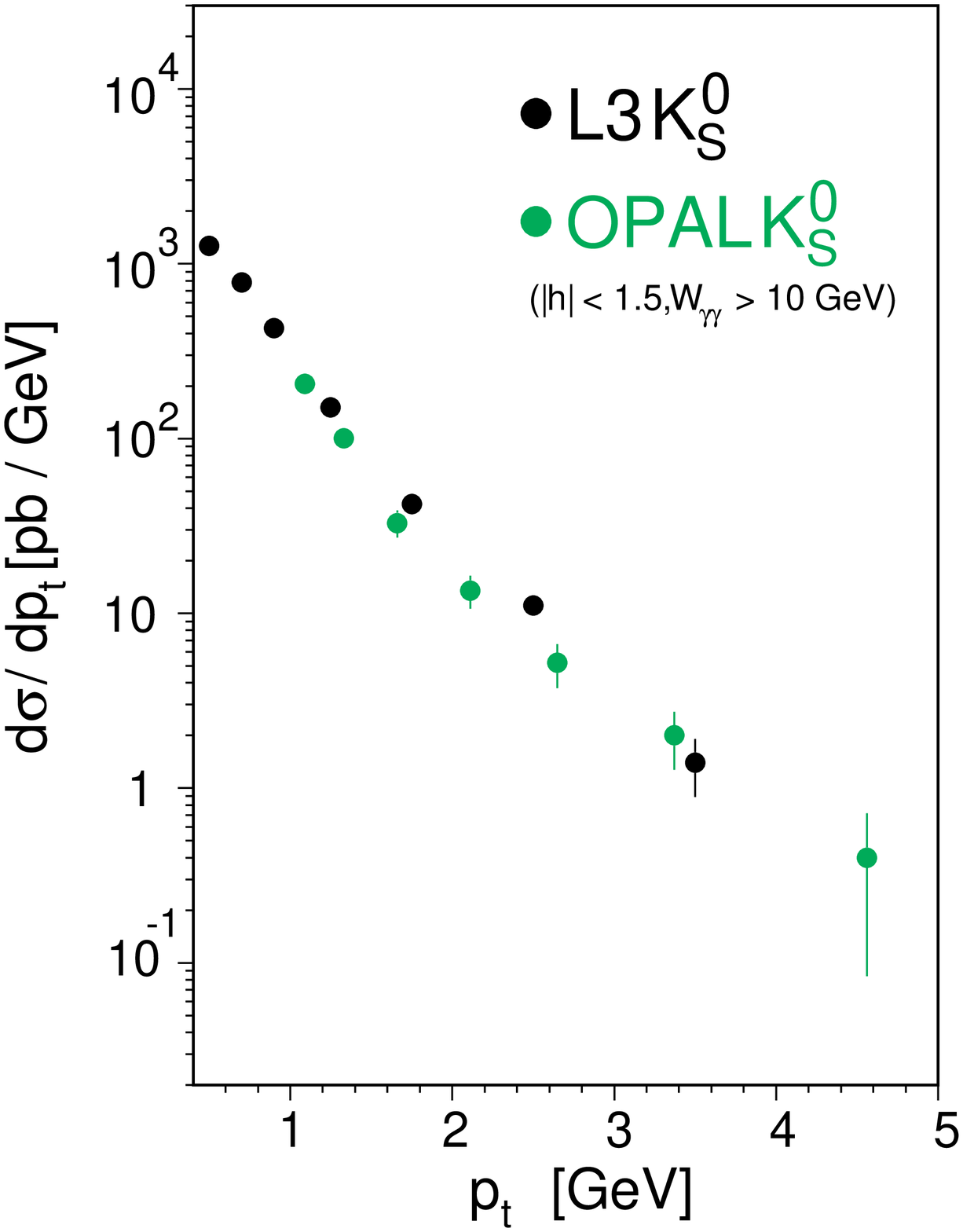,height=2.2in, width=2.1in}
    \psfig{figure=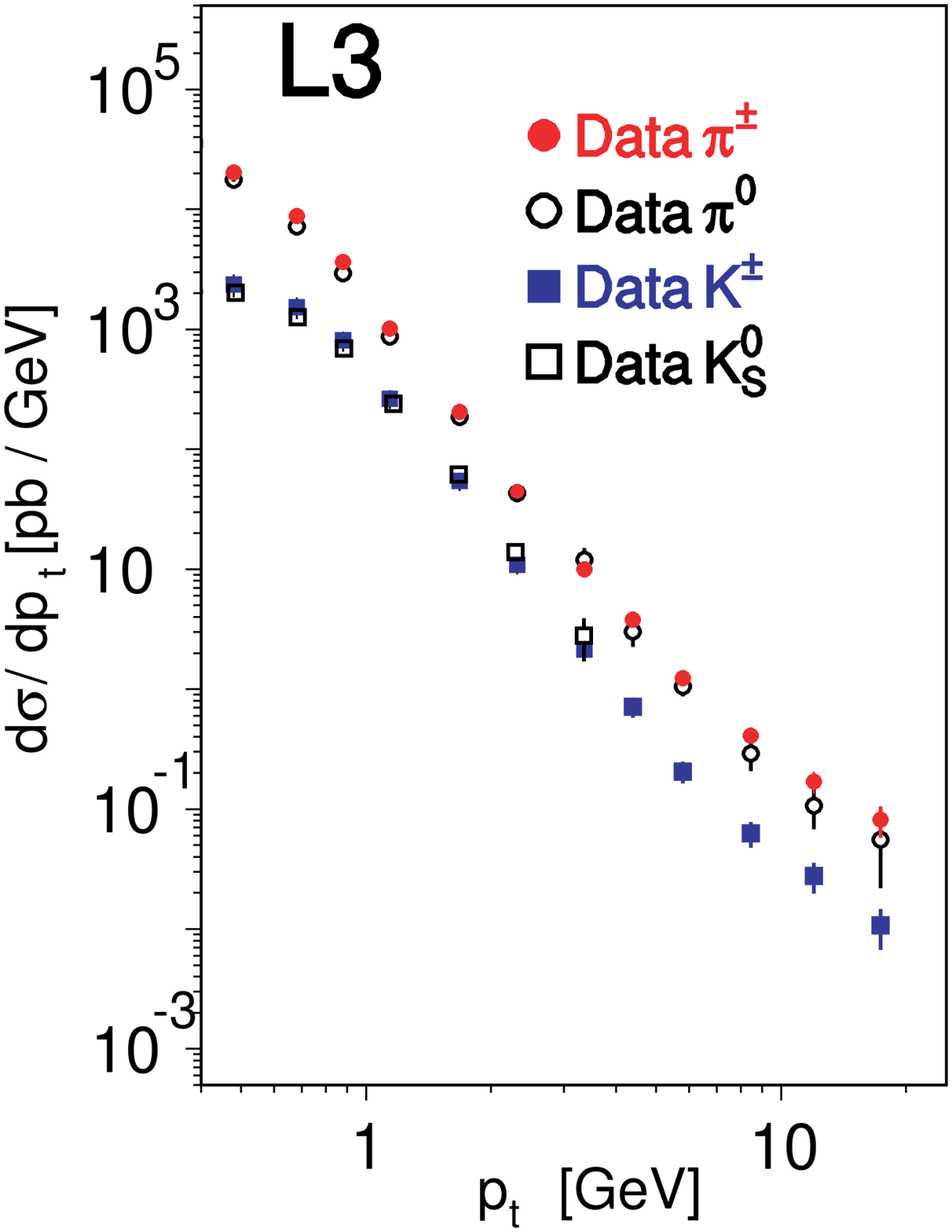,height=2.2in, width=2in}
    \caption{Differential cross-section $d\sigma/dp_t$ for inclusive charged
      particle ($\pi^\pm$ and $K^\pm$), $\pi^0$ and $K_S^0$ production.
      \label{fig:inp}}
  \end{center}
\end{figure}
\begin{figure}
  \begin{center}
    \vskip 0.5cm
    \psfig{figure=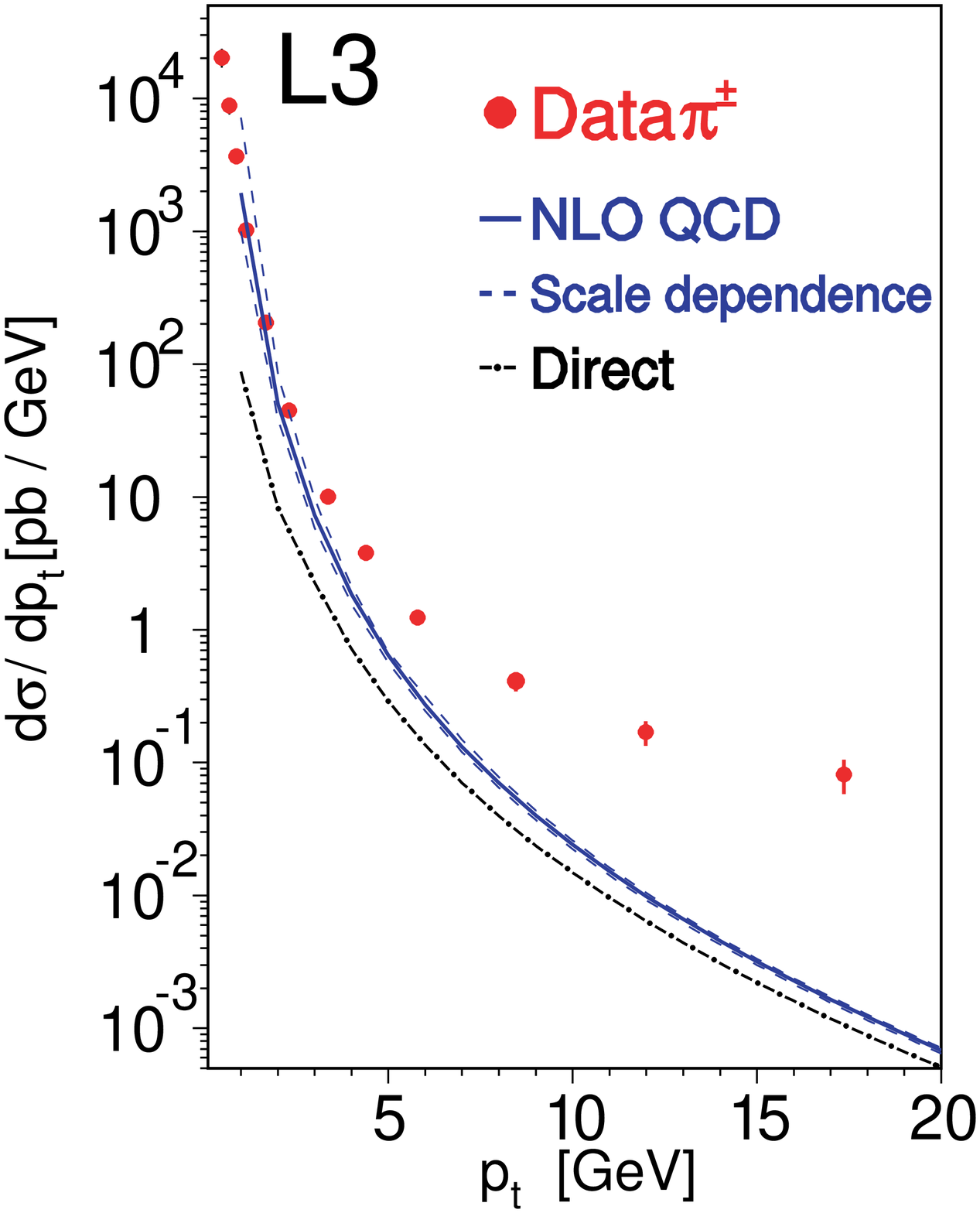,height=2.5in, width=2.6in}
    \psfig{figure=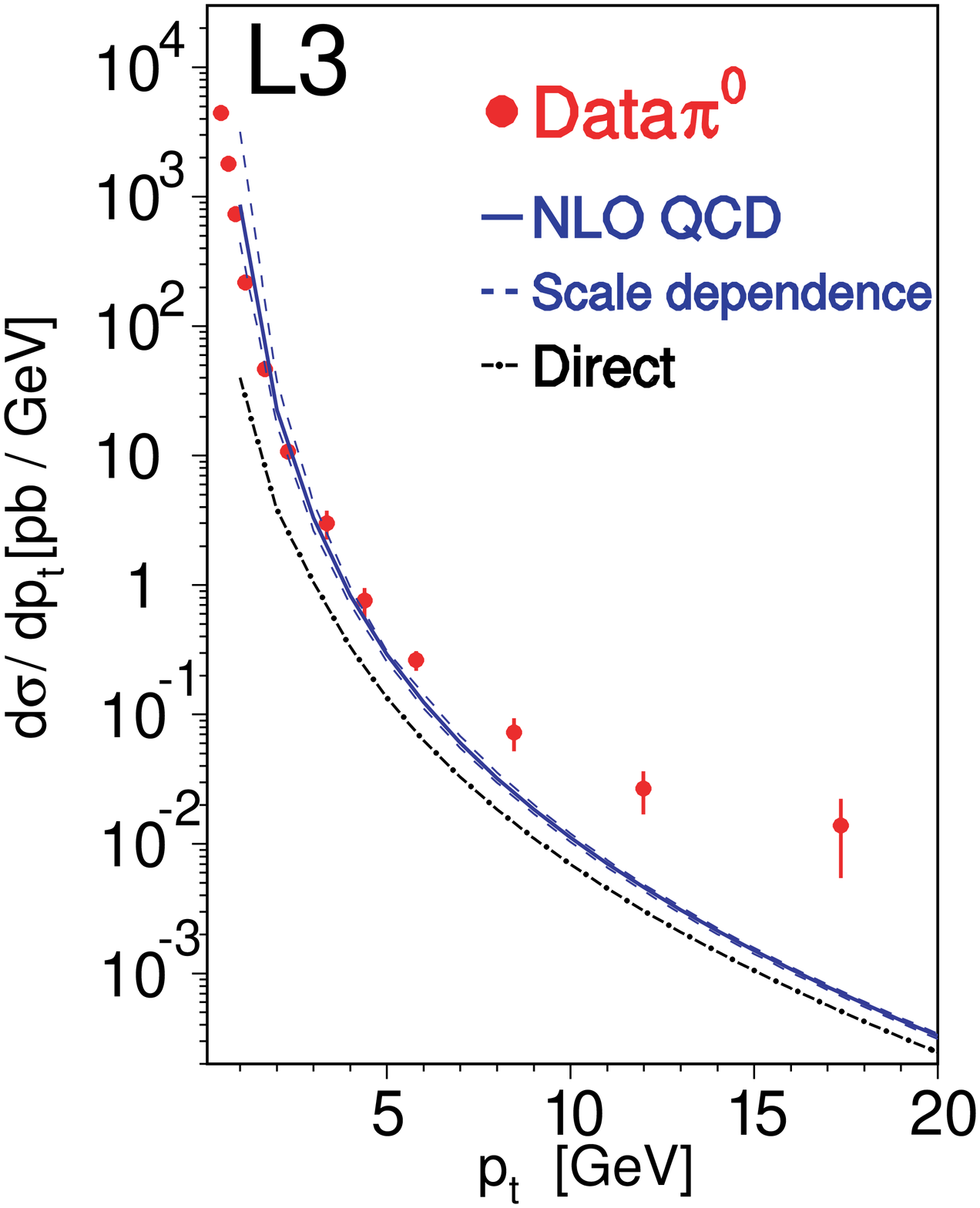,height=2.5in, width=2.6in}
    \caption{Inclusive $\pi^\pm$ and $\pi^0$ differential cross-section 
      $d\sigma/dp_t$ compared to NLO QCD calculations.
      \label{fig:nlo}}
  \end{center}
\end{figure}

Inclusive hadron production production in two-photon collisions can be used to
study the structure of photon interactions. The measurements~\cite{inc} were performed
at LEP at $\sqrt{s}=161 - 202~\mathrm{GeV}$. Several types of hadron, $\pi^\pm$,
$\pi^0$, $K^\pm$ and $K_S^0$, were measured or deduced from Monte Carlo fragmentation
functions~\cite{pythia}. Figure~\ref{fig:inp} 
shows the differential cross-section $d\sigma/dp_t$ for inclusive charged
particle ($\pi^\pm$ and $K^\pm$), $\pi^0$ and $K_S^0$ production. The results
from different experiments, L3 and OPAL, agree well. The measurement of charged
particles is also consistent with that of neutral particles after phase space renormalization.
This agreement tests Monte Carlo fragmentation functions.
In Figure~\ref{fig:nlo} the data are compared to 
analytical NLO QCD predictions~\cite{nlo}, which take into account both 
transverse and longitudinal virtual photons. The scale uncertainty in the NLO
calculation is also shown. The prediction is compatible with the data for
$p_t < 5~\mathrm{GeV}$. However, the distribution of the data is much flatter
than the NLO predictions at high $p_t$.

\section*{Acknowledgments}
I would like to acknowledge the LEP collaborations for providing me with
latest results of their analysis. I would like also to thank P. Achard,
A. Boehrer, B. Echenard, J.H. Field and M.N. Kienzle-Forccci for all their 
constructive suggestions.

\end{document}